\documentclass[aps,twocolumn,prl]{revtex4}
\usepackage{graphicx}
\usepackage{amssymb} 
\usepackage{url}     
\usepackage{bm}      
\usepackage{color}

\begin{document}

\title{Quantum Hall effect and Landau level crossing of Dirac fermions in trilayer graphene}
\author{Thiti  Taychatanapat$^{1}$}
\author{Kenji Watanabe$^{2}$}
\author{Takashi Taniguchi$^{2}$}
\author{Pablo Jarillo-Herrero$^{3}$}
\email{pjarillo@mit.edu}
\affiliation{$^{1}$Department of Physics, Harvard University, Cambridge, MA 02138 USA}
\affiliation{$^{2}$National Institute for Materials Science, Namiki 1-1, Tsukuba, Ibaraki 305-0044, Japan}
\affiliation{$^{3}$Department of Physics, Massachusetts Institute of Technology, Cambridge, MA 02139 USA}
\date{\today}
\begin{abstract}

We investigate electronic transport in high mobility (\textgreater 100,000 cm$^2$/V$\cdot$s) trilayer graphene (TLG) devices on hexagonal boron nitride, which enables the observation of Shubnikov-de Haas oscillations and an unconventional quantum Hall effect. The massless and massive characters of the TLG subbands lead to a set of Landau level crossings, whose magnetic field and filling factor coordinates enable the direct determination of the Slonczewski-Weiss-McClure (SWMcC) parameters used to describe the peculiar electronic structure of trilayer graphene. Moreover, at high magnetic fields, the degenerate crossing points split into manifolds indicating the existence of broken-symmetry quantum Hall states.

\end{abstract}

\maketitle

Bernal or ABA stacked TLG (Fig.~1b) is an intriguing material to study Dirac physics and quantum Hall effect (QHE) because of its unique band structure which, in the simplest approximation, consists of massless single-layer graphene (SLG) and massive bilayer graphene (BLG) subbands at low energy (Fig. 1c)\cite{Lu_ABAbandStructure,Guinea_ABAbandStructure,Latil_ABAbandStructure,Partoens_ABAbandStructure}.  The energies of the Landau levels (LLs)  for massless charge carriers depend on the square root of the magnetic field $\sqrt{B}$~\cite{Novoselov_QHE_single,Zhang_QHE,Li_STS,Miller_TMCO} while for massive charge carriers they depend linearly on $B$~\cite{Novoselov_QHE_bi,McCann_BiQHE,Li_STS}. Therefore, the LLs from these two different subbands in TLG should cross at some finite fields, resulting in accidental LL degeneracies at the crossing points.

However, one of the major challenges so far to observe QHE in TLG has been its low mobility on SiO$_2$ substrates\cite{Craciun_TrilayerGate,Zhu_mobilityTri}. To overcome this problem, we use high quality hexagonal boron nitride (hBN) single crystals\cite{Taniguchi_BN} as local substrates, which have been shown to reduce carrier scattering in graphene devices\cite{Dean_BN}. Substrate supported devices also allows us to reach higher carrier density than suspended samples~\cite{Bao_suspendedTri}, which is necessary for the observation of the LL crossings.

\begin{figure}[!h]
\begin{center}
\includegraphics[width = 3in]{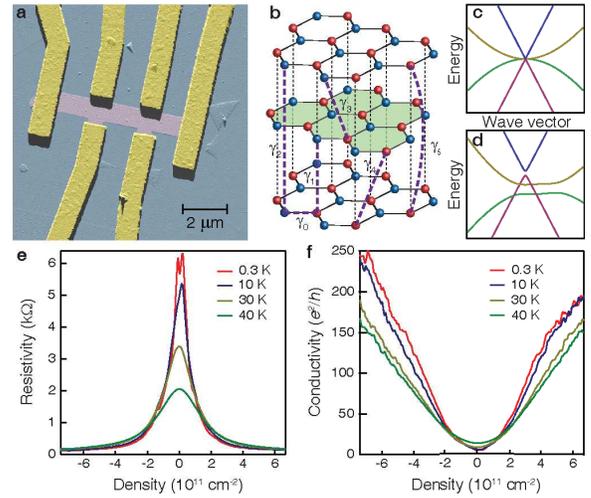}
\caption{\textbf{a}) False color atomic force microscopy image of a TLG Hall bar device on hBN. \textbf{b}) Bernal stacked TLG atomic lattice. The SWMcC hopping parameters, $\gamma_i$, are shown by purple dashed lines connecting the corresponding hopping sites. In addition to $\gamma_i$, the SWMcC parameters also include the on-site energy difference, $\delta$, between A and B sublattices (blue and red lattices). \textbf{c}) Band structure of TLG at low energy, which takes into account only the nearest neighbour intra- and inter-layer hopping parameters $\gamma_0$ and $\gamma_1$. \textbf{d}) Band structure of TLG within a full parameter model, with the parameters calculated from the SdH oscillations in Fig. 2b. \textbf{e}) Resistivity as a function of density and temperature for TLG. The double peak structure starts to emerge as temperature decreases below 10 K.  \textbf{f}) Conductivity as a function of density and temperature.The field-effect mobility at 300 mK reaches $\sim$110,000 cm$^2$/V$\cdot$s and decreases to $\sim$65,000 cm$^2$/V$\cdot$s at 40 K} \label{F:Fig1}
\end{center}
\end{figure}

Figure 1a shows an atomic force microscope image of a Hall bar shaped TLG device on hBN. Our fabrication process consists of mechanically exfoliating hBN and graphene flakes on different supports, and a flip chip bonding step to align them on top of each other (see SI for details). The graphene flakes are then patterned into a Hall bar geometry and contacted by electron beam lithography. The device is then annealed in forming gas to remove residue and cooled down in a He-3 cryostat.

\begin{figure*}
\begin{center}
\includegraphics{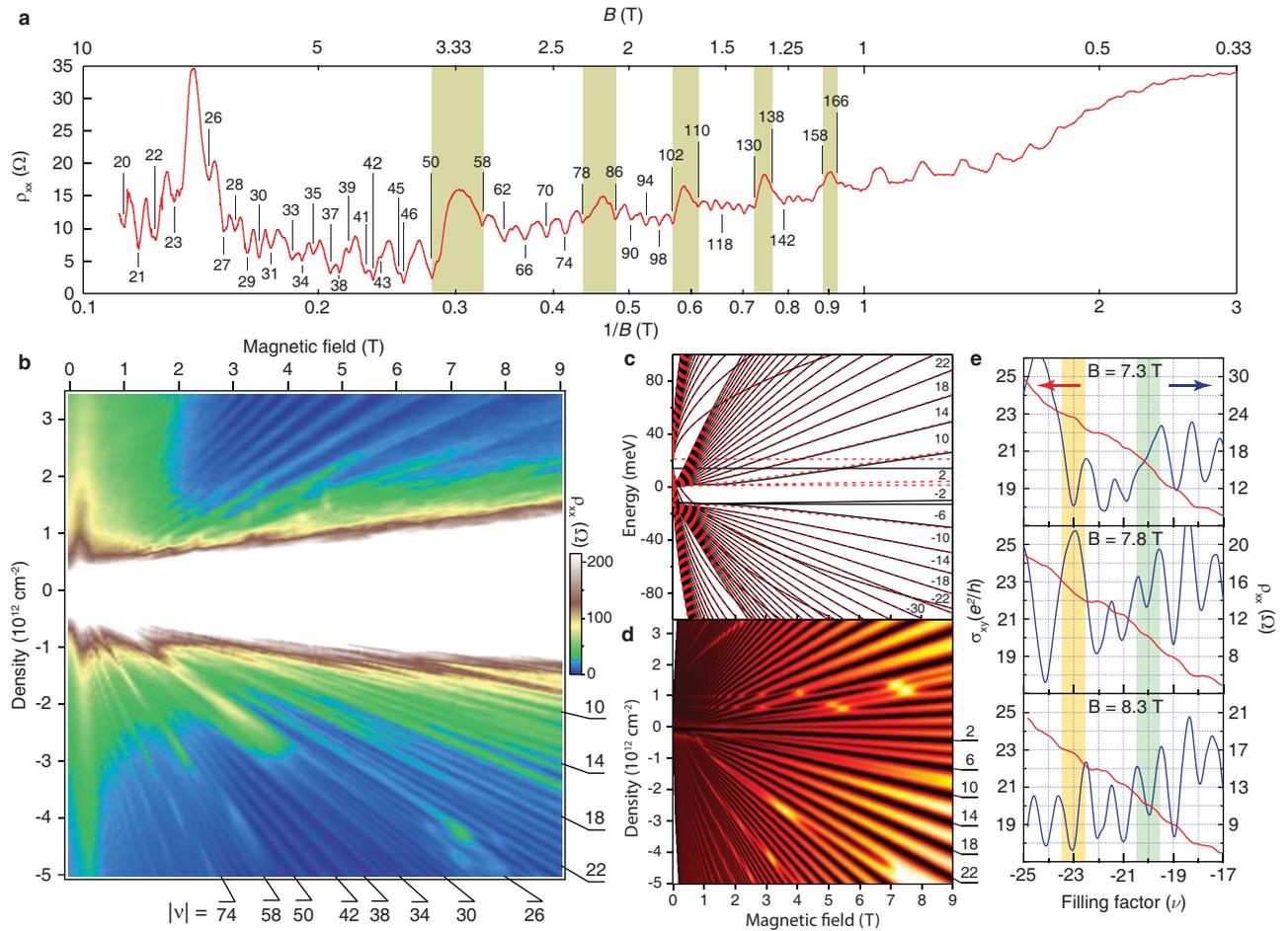}
\caption{\textbf{a}) $\rho_{\mathrm{xx}}$ as a function of inverse magnetic field at 300 mK. The numbers inside the figure indicate the filling factors at the SdH oscillation minima. The highlighted bands show the regions of 8-fold degeneracy, which provide evidence for LL crossings of the SLG- and BLG-like subbands. For $B > 4$ T, the SdH minima are separated by $\Delta\nu = 1$ or $2$, indicating the splitting of LLs. \textbf{b}) Color map of $\rho_{\mathrm{xx}}$ versus $n$ and $B$ at 300 mK. The diagonal lines correspond to constant filling factor lines. The beating pattern, most visible at negative densities, is a consequence of LL crossings. The white central region corresponds to an insulating state at zero density (see SI). \textbf{c}) Calculated LL energy spectrum in TLG for the SWMcC parameters obtained from \textbf{b}. The red dashed and black lines are LLs at K and K$'$ points respectively. The roughly $\sqrt{B}$-like and linear $B$-like dispersion from the SLG- and BLG-like subbands is evident. Each line corresponds to a spin degenerate LL. \textbf{d}) Calculated density of states as a function of density and $B$ from the LL spectrum in \textbf{c}. Apart from the LL splitting, the location of the LL crossings agrees very well with the experimental data in \textbf{b}.  \textbf{e}) $\rho_{\mathrm{xx}}$ and $\sigma_{\mathrm{xy}}$ as a function of filling factor for $B = 7.3, 7.8,$ and $8.3$ T. The highlighted orange region shows the appearance of the LL crossing at $\nu = -23$ while the green highlighted region shows the LL splitting occurring at $\nu = -20$.} \label{F:Fig2}
\end{center}
\end{figure*}

In order to further reduce disorder and increase the mobility, we perform current annealing at low temperature\cite{Moser_CurrentAnneal}. Figures 1e and 1f show the resistivity and conductivity of a TLG device at zero magnetic field after current annealing. The resistivity at the Dirac peak exhibits a strong temperature dependence, which in SLG is a strong indication of high device quality\cite{Du_suspended,Bolotin_Tdepend}. In addition, we also observe a double-peak structure at low temperatures (Fig. 1e). This double-peak structure is likely due to the band overlap which occurs in TLG when all SWMcC parameters are included in the tight-binding calculation of its band structure, as we show below. The field effect mobility of this device reaches $110,000$~cm$^2$/V$\cdot$s at $300$ mK at densities as high as $6\times 10^{11}\,$cm$^{-2}$ . This mobility value is two orders of magnitude higher than previously reported values for supported TLG\cite{Craciun_TrilayerGate,Zhu_mobilityTri} and comparable to suspended SLG-TLG samples\cite{Bolotin_Tdepend,Bao_suspendedTri}. The low disorder and high mobility enable us to probe LL crossings of Dirac fermions through the measurement of Shubnikov-de Hass (SdH) oscillations.

Figure 2a shows longitudinal resistivity $\rho_{\mathrm{xx}}$ as a function of $1/B$, for a carrier density $n=-4.4 \times 10^{12}\,$cm$^{-2}$. A pattern of SdH oscillations is clearly visible, albeit with different visibility and features depending on the $B$ range. At low $B$ (below $\sim1$~T), there are a number of oscillations characterized by broad minima separated by relatively narrower maxima. Beyond $\sim1$~T, the minima become sets of narrower oscillations, and a clear pattern emerges: each minimum in the oscillations indicates a completely filled LL with corresponding filling factor $\nu = hn/eB$, where $h$ is Planck's constant, and $e$ is the electron charge. Within a single particle picture, each LL is 4-fold degenerate, the degeneracy originating from the valley (K and K$'$) and spin (up and down) degrees of freedoms in both the SLG-like and BLG-like subbands. When LLs from these two subbands cross at a given $B$, the coexistence of two 4-fold degenerate LLs increases the degeneracy to 8-fold. This 8-fold degeneracy is highlighted by the green bands in Fig. 2a, where $\nu$ changes by 8 from minimum to minimum instead of by 4. For $B\geq4$~T, the splitting of the LLs results in $\nu$ changing by either 1 or 2, as the different broken-symmetry quantum Hall states are occupied.

A more complete understanding of the TLG LL energy spectrum is obtained by plotting $\rho_{\mathrm{xx}}$ as a function of $n$ and $B$ as shown in Figure 2b. The resulting fan diagram lines correspond to the SdH oscillations mentioned above, while the white central region corresponds to an insulating behavior at $\nu=0$ (see SI for details).  The abovementioned crossings of SLG-like and BLG-like LLs manifest themselves as a beating pattern in the SdH oscillations, with a greater number of them and more visible on the hole side ($n<0$). This electron-hole asymmetry results from the TLG band structure, as we show below. In addition, the LL splittings appear as finer split lines in the SdH oscillations. For each LL crossing, there is an enhancement of $\rho_{\mathrm{xx}}$ due to the enhanced density of states\cite{Piazza_PhaseTranQHF,Zhang_LLcrossingTwo-subband}, and each crossing point can be uniquely identified by $B$ and $\nu$. For instance, at $B \sim 3\,$T and $n \sim -4\times10^{12}\,$cm$^{-2}$, the filling factors associated with the minima in the corresponding SdH oscillations change from 50 to 58 indicating that the crossing occurs at $\nu = 54$.

The positions of the crossings in $B$ and $\nu$ space depend sensitively on the TLG band structure, and therefore enable a direct electronic transport determination of the relevant SWMcC parameters for TLG. These parameters, proposed to explain the band structure of graphite\cite{Dresselhaus_Graphite}, describe the different intra- and inter-layer hopping terms in the different graphene sheets (Fig. 1b). We note that TLG is the fewest layer graphene system whose description includes all the SWMcC parameters. The simplest TLG model, in which only the nearest intra- and inter-layer couplings ($\gamma_0$ and $\gamma_1$) are considered (the ones typically used to describe SLG and BLG), results in symmetric electron and hole bands (Fig. 1c) and therefore is clearly insufficient to explain the experimental data. We therefore use all the relevant SWMcC parameters to numerically calculate the LL energy spectrum (Fig. 2c) and density of states as a function of $B$ (Fig. 2d), and perform a minimization procedure to fit the experimental data in Fig. 2b. In order to lower the number of parameters, we take $\gamma_0 = 3.1$ eV, $\gamma_1 = 0.39$ eV and $\gamma_3 = 0.315$ eV (see SI), and we obtain from our fit the following values of the SWMcC parameters; $\gamma_2 = -0.028(4)$ eV, $\gamma_4 = 0.041(10)$ eV, $\gamma_5 = 0.05(2)$ eV, and $\delta = 0.046(10)$ eV. The definitions of the $\gamma_i$ can be found in Fig. 1b and $\delta$ is the on-site energy difference between the two-inequivalent carbon sublattices residing in the same graphene layer. The values of the SWMcC parameters obtained are similar to previously reported values for graphite\cite{Dresselhaus_Graphite} and, apart from the broken-symmetry states (see discussion below), our data agree very well with the LLs corresponding to Bernal stacked TLG, and not to rhombohedral stacked TLG\cite{Mikito_ABC}. These parameters result in the overall electron-hole asymmetric band structure shown in Fig. 1d, with small band gaps $E_{\mathrm{g,S}} \sim 7$~meV and $E_{\mathrm{g,B}}\sim 14$~meV, for the SLG- and BLG-like subbands, and a band overlap $E_{\mathrm{o}}\sim 14$~meV.

\begin{figure}
\begin{center}
\includegraphics[width = 3in]{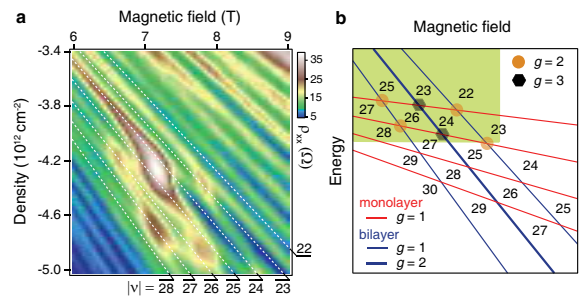}
\caption{\textbf{a}) $\rho_{\mathrm{xx}}$ as a function of density and $B$ at 300 mK showing a manifold of LL crossing points. The high $\rho_{\mathrm{xx}}$ regions correspond to enhanced degeneracy due to LL crossings. Five crossing points are clearly visible and the sixth point is starting to appear in the lower-right corner. White dashed lines are guides to the eye for each $\nu$ labeled on the edges. \textbf{b}) Schematic splitting and crossing of LLs yielding the manifold of crossings shown in \textbf{a}. Red and blue lines represent the split LL spectrum for the broken-symmetry QH states of the $N = -1$ LL from the SLG-like subband and the $N = -5$ LL from the BLG-like subband, respectively. The degeneracies for each level are $g=1$ for thin lines and $g=2$ for the thick line. The highlighted green area corresponds to the region observed in the data in \textbf{a}. The numbers inside each region show the corresponding filling factors} \label{F:Fig4}
\end{center}
\end{figure}

The LLs in TLG are not truly 4-fold degenerate even in a single particle picture, owing to the finite value of $\gamma_2$, $\gamma_5$, and $\delta$, which break valley degeneracy (see Fig. 2c), in addition to the Zeeman interaction which breaks spin degeneracy. Our data at high $B$ (Fig. 2a and 2b) show that the splitting of 4-fold degenerate LLs is observed up to filling factors as high as $\nu=46$. While single particle effects may partly explain these broken-symmetry QH states (e.g. from the width of the LLs crossings, we estimate the disorder broadening of the LLs to be $\sim$1 mV  which is similar to the the Zeeman splitting at $\sim$8 T), it is likely that electron-electron (e-e) interactions play a significant role too, as it is the case in SLG and BLG\cite{Zhang_MonoLLSplitting,Checkelsky_nuZero,Feldman_bilayer,Zhao_SymBreakBi,Dean_BN}. For example, the insulating behavior we observe at $\nu=0$, cannot be explained by single particle effects, given the band overlap between the SLG- and BLG-like subbands, and the single particle LL energy spectrum shown in Fig. 2c. However, a more detailed study including measurements of the gap energies and measurements in tilted magnetic fields, beyond the scope of this paper, is necessary to investigate the precise role of e-e interactions in TLG. Figure 2e shows example traces where the different behavior of LL crossings and LL splitting  can be seen.

At high $B$, the LL crossing points should become crossing manifolds due to the crossing between the split SLG- and BLG-like LLs. One such example is shown in Fig. 3a. From the LL energy spectrum shown in Fig 2c, the manifold corresponds to the crossing between the $N = -1$ LL of the SLG-like subband, LL$_\mathrm{S}^{-1}$, and the $N = -5$ LL of the BLG-like subband, LL$_\mathrm{B}^{-5}$. In order to reproduce the observed degeneracies at the crossings, the 4-fold LL$_\mathrm{S}^{-1}$ has to completely split into four singly-degenerate LLs while the 4-fold LL$_\mathrm{B}^{-5}$ splits into 3 LLs: two singly degenerate LLs and one doubly degenerate LL. Figure 3b shows schematically the full 12-point manifold, of which only 6 crossing points are visible in our density and $B$ range. We have found that this splitting scheme is the only one that yields the correct result for both the degeneracies at the crossings and the filling factors at which they occur. The observation of the full 4-fold splitting of the LL$_\mathrm{S}^{-1}$ in TLG, although expected, is remarkable since previous transport studies of the $N=1$ LL in SLG had reported only the breaking of some of the degeneracies\cite{Zhang_MonoLLSplitting,Du_fraction}, and the full 4-fold splitting has only been seen in recent STM experiments\cite{Song_N1splitting}. The 1-2-1 splitting of LLs from the BLG-like subband, however, is more anomalous. Naively, one would expect the splitting to be either 2 fold or 4 fold, depending on whether one of the two degrees of freedom (valley or spin) is split or both are\cite{Feldman_bilayer,Zhao_SymBreakBi}. However, we note that this 1-2-1 splitting may also be present in a recent study of BLG on hBN in the intermediate $B$-regime\cite{Dean_BN}, and may possibly indicate a richer phase diagram based on SU(4) rather than SU(2)xSU(2) symmetry breaking. A detailed study of the crossing between spin/valley polarized LLs of massless and massive Dirac Fermions, together with the aforementioned possible role of e-e interactions, could potentially lead to some intriguing phenomena such as phase transitions in quantum Hall ferromagnets\cite{Jungwirth_QHF,Piazza_PhaseTranQHF}.

\begin{figure}
\begin{center}
\includegraphics[width = 2.5 in]{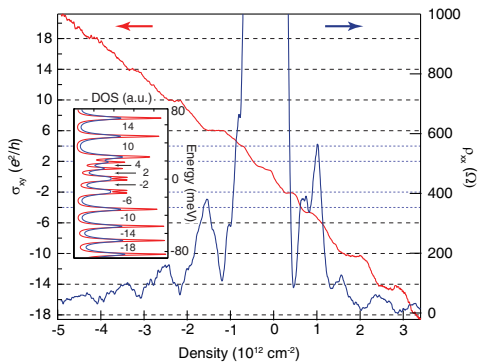}
\caption{$\sigma_{\mathrm{xy}}$ and $\rho_{\mathrm{xx}}$ as a function of density at $B = 9$ T and $T = 300$ mK, and before the last current annealing step. The dashed lines indicate the expected QH plateaus based on the simplest TLG model approximation. The dotted lines indicate the extra QH plateaus based on the full band structure determined from Fig. 2c. (Inset) Calculated Density of states using full SWMcC parameter model. The blue line is calculated using higher disorder broadening than the red line.} \label{F:Fig3}
\end{center}
\end{figure}

Although the splitting of the LLs at high $B$ provides insight into broken symmetries in TLG in the QH regime, it also masks out the QH plateaus expected within the simplest single particle model for TLG. The sequence of plateaus arising from such simple models has proven a useful tool in identifying SLG and BLG\cite{Novoselov_QHE_single,Zhang_QHE,Novoselov_QHE_bi}.
For completeness, Figure 4 shows $\rho_{\mathrm{xx}}$ and $\sigma_{\mathrm{xy}}$ at $B = 9$ T before current annealing, i.e. in the presence of increased disorder which prevents the observation of LL splitting. In the simplest model, the QHE plateaus are expected at $\sigma_{\mathrm{xy}}=\pm4(N+1/2+1)e^2/h$ for $N = 0, 1, \ldots$ where the 12-fold zero energy LL results from the 4-fold and 8-fold zero energy LLs of the SLG- and BLG-like subbands, respectively\cite{Motohiko_QHEtrilayer,Mikito_ParityValley}. Our observations agree with this simple prediction for $|\nu| \geq 10$ (with observed plateaus at $\pm10, \pm14, \pm18 e^2/h$), but we observe in addition extra plateaus for $\nu = \pm2$ and $\pm4$ as well as the absence of a plateau at $\nu = +6$. This unconventional QHE can be explained within the band model calculated using the SWMcC parameters obtained from Fig. 2a-c. In such model, the non-zero values of $\gamma_2$, $\gamma_5$, and $\delta$ lift the degeneracy of the ``zero-energy" LLs of the SLG- and BLG-like subbands (Fig. 2c).  In addition, the 4-fold degenerate $N = 0$ LL of the SLG-like subband splits into two 2-fold degenerate valley polarized LLs and the 8-fold degenerate (spin, valley and N=0,1 LLs) zero energy LLs of the BLG-like subband splits into two 4-fold degenerate LLs (the splitting between $N=0$ and $N=1$ LLs remains relatively small compared to the valley splitting). We note that the Zeeman splitting is at least an order of magnitude smaller than other types of splitting even at 9 T which is the reason why LLs remain spin degenerate in this non-interacting model.

The inset to Fig. 4 shows the calculated density of states as a function of energy at 9 T. The zero density is located between two nearly degenerate LLs, each with 2-fold degeneracy which explains the observed plateaus at $\nu = \pm2$. The absence of a plateau at $\nu = 0$ is likely due to disorder, which smears out the small energy gap between these two LLs. The plateaus at $\nu = \pm4$ stem from the next 2-fold degenerate LLs. However, these plateaus are not yet completely developed at $9$~T, especially the one at $\nu = -4$ ($\sigma_{\mathrm{xy}}= 4e^2/h$) which coincides with the small energy gap between this LL and the next one. Finally, the absence of a plateau at $\nu = +6$ ($\sigma_{\mathrm{xy}}=-6e^2/h$) is due to the crossing between a 2-fold and a 4-fold degenerate LL. The degeneracy at the crossing becomes 6-fold and causes the position of the plateau to step from $\nu = 4$ to $\nu = 10$ (the non-developed $\nu = 4$ plateau does not reach its exact value at $\sigma_{\mathrm{xy}}=-4e^2/h$). Unlike SLG and BLG in which the sequence of the plateaus are the same for all $B$, the observed plateaus in TLG depend on $B$ because of the LL crossing.

We thank M. Koshino and E. McCann for discussions and sharing their preliminary work on LLs in Bernal stacked TLG. We also thank L. Levitov and P. Kim for discussions, A. F. Young for discussions and experimental help on hBN, and J. D. Sanchez-Yamagishi and J. Wang for experimental help. We acknowledge financial support from the Office of Naval Research GATE MURI and a National Science Foundation Career Award. This reserach has made use of the NSF funded MIT CMSE and Harvard CNS facilities.


\begin{thebibliography}{31}
\expandafter\ifx\csname natexlab\endcsname\relax\def\natexlab#1{#1}\fi
\expandafter\ifx\csname bibnamefont\endcsname\relax
  \def\bibnamefont#1{#1}\fi
\expandafter\ifx\csname bibfnamefont\endcsname\relax
  \def\bibfnamefont#1{#1}\fi
\expandafter\ifx\csname citenamefont\endcsname\relax
  \def\citenamefont#1{#1}\fi
\expandafter\ifx\csname url\endcsname\relax
  \def\url#1{\texttt{#1}}\fi
\expandafter\ifx\csname urlprefix\endcsname\relax\def\urlprefix{URL }\fi
\providecommand{\bibinfo}[2]{#2}
\providecommand{\eprint}[2][]{\url{#2}}

\bibitem[{\citenamefont{Lu et~al.}(2006)\citenamefont{Lu, Chang, Huang, Chen,
  and Lin}}]{Lu_ABAbandStructure}
\bibinfo{author}{\bibfnamefont{C.~L.} \bibnamefont{Lu}},
  \bibinfo{author}{\bibfnamefont{C.~P.} \bibnamefont{Chang}},
  \bibinfo{author}{\bibfnamefont{Y.~C.} \bibnamefont{Huang}},
  \bibinfo{author}{\bibfnamefont{R.~B.} \bibnamefont{Chen}}, \bibnamefont{and}
  \bibinfo{author}{\bibfnamefont{M.~L.} \bibnamefont{Lin}},
  \bibinfo{journal}{Phys. Rev. B} \textbf{\bibinfo{volume}{73}},
  \bibinfo{pages}{144427} (\bibinfo{year}{2006}).

\bibitem[{\citenamefont{Guinea et~al.}(2006)\citenamefont{Guinea, Neto, and
  Peres}}]{Guinea_ABAbandStructure}
\bibinfo{author}{\bibfnamefont{F.}~\bibnamefont{Guinea}},
  \bibinfo{author}{\bibfnamefont{A.~H.~C.} \bibnamefont{Neto}},
  \bibnamefont{and} \bibinfo{author}{\bibfnamefont{N.~M.~R.}
  \bibnamefont{Peres}}, \bibinfo{journal}{Phys. Rev. B}
  \textbf{\bibinfo{volume}{73}}, \bibinfo{pages}{245426}
  (\bibinfo{year}{2006}).

\bibitem[{\citenamefont{Latil and Henrard}(2006)}]{Latil_ABAbandStructure}
\bibinfo{author}{\bibfnamefont{S.}~\bibnamefont{Latil}} \bibnamefont{and}
  \bibinfo{author}{\bibfnamefont{L.}~\bibnamefont{Henrard}},
  \bibinfo{journal}{Phys. Rev. Lett.} \textbf{\bibinfo{volume}{97}},
  \bibinfo{pages}{036803} (\bibinfo{year}{2006}).

\bibitem[{\citenamefont{Partoens and
  Peeters}(2006)}]{Partoens_ABAbandStructure}
\bibinfo{author}{\bibfnamefont{B.}~\bibnamefont{Partoens}} \bibnamefont{and}
  \bibinfo{author}{\bibfnamefont{F.~M.} \bibnamefont{Peeters}},
  \bibinfo{journal}{Phys. Rev. B} \textbf{\bibinfo{volume}{74}},
  \bibinfo{pages}{075404} (\bibinfo{year}{2006}).

\bibitem[{\citenamefont{Novoselov et~al.}(2005)\citenamefont{Novoselov, Geim,
  Morozov, Jiang, Katsnelson, Grigorieva, Dubonos, and
  Firsov}}]{Novoselov_QHE_single}
\bibinfo{author}{\bibfnamefont{K.~S.} \bibnamefont{Novoselov}},
  \bibinfo{author}{\bibfnamefont{A.~K.} \bibnamefont{Geim}},
  \bibinfo{author}{\bibfnamefont{S.~V.} \bibnamefont{Morozov}},
  \bibinfo{author}{\bibfnamefont{D.}~\bibnamefont{Jiang}},
  \bibinfo{author}{\bibfnamefont{M.~I.} \bibnamefont{Katsnelson}},
  \bibinfo{author}{\bibfnamefont{I.~V.} \bibnamefont{Grigorieva}},
  \bibinfo{author}{\bibfnamefont{S.~V.} \bibnamefont{Dubonos}},
  \bibnamefont{and} \bibinfo{author}{\bibfnamefont{A.~A.}
  \bibnamefont{Firsov}}, \bibinfo{journal}{Nature}
  \textbf{\bibinfo{volume}{438}}, \bibinfo{pages}{197} (\bibinfo{year}{2005}).

\bibitem[{\citenamefont{Zhang et~al.}(2005{\natexlab{a}})\citenamefont{Zhang,
  Tan, Stormer, and Kim}}]{Zhang_QHE}
\bibinfo{author}{\bibfnamefont{Y.}~\bibnamefont{Zhang}},
  \bibinfo{author}{\bibfnamefont{Y.-W.} \bibnamefont{Tan}},
  \bibinfo{author}{\bibfnamefont{H.~L.} \bibnamefont{Stormer}},
  \bibnamefont{and} \bibinfo{author}{\bibfnamefont{P.}~\bibnamefont{Kim}},
  \bibinfo{journal}{Nature} \textbf{\bibinfo{volume}{438}},
  \bibinfo{pages}{201} (\bibinfo{year}{2005}{\natexlab{a}}).

\bibitem[{\citenamefont{Li and Andrei}(2007)}]{Li_STS}
\bibinfo{author}{\bibfnamefont{G.}~\bibnamefont{Li}} \bibnamefont{and}
  \bibinfo{author}{\bibfnamefont{E.~Y.} \bibnamefont{Andrei}},
  \bibinfo{journal}{Nature Phys.} \textbf{\bibinfo{volume}{3}},
  \bibinfo{pages}{623} (\bibinfo{year}{2007}).

\bibitem[{\citenamefont{Miller et~al.}(2009)\citenamefont{Miller, Kubista,
  Rutter, Ruan, de~Heer, First, and Stroscio}}]{Miller_TMCO}
\bibinfo{author}{\bibfnamefont{D.~L.} \bibnamefont{Miller}},
  \bibinfo{author}{\bibfnamefont{K.~D.} \bibnamefont{Kubista}},
  \bibinfo{author}{\bibfnamefont{G.~M.} \bibnamefont{Rutter}},
  \bibinfo{author}{\bibfnamefont{M.}~\bibnamefont{Ruan}},
  \bibinfo{author}{\bibfnamefont{W.~A.} \bibnamefont{de~Heer}},
  \bibinfo{author}{\bibfnamefont{P.~N.} \bibnamefont{First}}, \bibnamefont{and}
  \bibinfo{author}{\bibfnamefont{J.~A.} \bibnamefont{Stroscio}},
  \bibinfo{journal}{Science} \textbf{\bibinfo{volume}{324}},
  \bibinfo{pages}{924} (\bibinfo{year}{2009}).

\bibitem[{\citenamefont{Novoselov et~al.}(2006)\citenamefont{Novoselov, McCann,
  Morozov, Fal'ko, Katsnelson, Zeitler, Jiang, schedin, and
  Geim}}]{Novoselov_QHE_bi}
\bibinfo{author}{\bibfnamefont{K.~S.} \bibnamefont{Novoselov}},
  \bibinfo{author}{\bibfnamefont{E.}~\bibnamefont{McCann}},
  \bibinfo{author}{\bibfnamefont{S.~V.} \bibnamefont{Morozov}},
  \bibinfo{author}{\bibfnamefont{V.~I.} \bibnamefont{Fal'ko}},
  \bibinfo{author}{\bibfnamefont{M.~I.} \bibnamefont{Katsnelson}},
  \bibinfo{author}{\bibfnamefont{U.}~\bibnamefont{Zeitler}},
  \bibinfo{author}{\bibfnamefont{D.}~\bibnamefont{Jiang}},
  \bibinfo{author}{\bibfnamefont{F.}~\bibnamefont{schedin}}, \bibnamefont{and}
  \bibinfo{author}{\bibfnamefont{A.~K.} \bibnamefont{Geim}},
  \bibinfo{journal}{Nature Phys.} \textbf{\bibinfo{volume}{2}},
  \bibinfo{pages}{177} (\bibinfo{year}{2006}).

\bibitem[{\citenamefont{McCann and Fal'ko}(2006)}]{McCann_BiQHE}
\bibinfo{author}{\bibfnamefont{E.}~\bibnamefont{McCann}} \bibnamefont{and}
  \bibinfo{author}{\bibfnamefont{V.~L.} \bibnamefont{Fal'ko}},
  \bibinfo{journal}{Phys. Rev. Lett.} \textbf{\bibinfo{volume}{96}},
  \bibinfo{pages}{086805} (\bibinfo{year}{2006}).

\bibitem[{\citenamefont{Craciun et~al.}(2009)\citenamefont{Craciun, Russo,
  Yamamoto, Oostinga, Morpurgo, and Tarucha}}]{Craciun_TrilayerGate}
\bibinfo{author}{\bibfnamefont{M.~F.} \bibnamefont{Craciun}},
  \bibinfo{author}{\bibfnamefont{S.}~\bibnamefont{Russo}},
  \bibinfo{author}{\bibfnamefont{M.}~\bibnamefont{Yamamoto}},
  \bibinfo{author}{\bibfnamefont{J.~B.} \bibnamefont{Oostinga}},
  \bibinfo{author}{\bibfnamefont{A.~F.} \bibnamefont{Morpurgo}},
  \bibnamefont{and} \bibinfo{author}{\bibfnamefont{S.}~\bibnamefont{Tarucha}},
  \bibinfo{journal}{Nature Nanotech.} \textbf{\bibinfo{volume}{4}},
  \bibinfo{pages}{383} (\bibinfo{year}{2009}).

\bibitem[{\citenamefont{Zhu et~al.}(2009)\citenamefont{Zhu, Perebeinos,
  Freitag, and Avouris}}]{Zhu_mobilityTri}
\bibinfo{author}{\bibfnamefont{W.}~\bibnamefont{Zhu}},
  \bibinfo{author}{\bibfnamefont{V.}~\bibnamefont{Perebeinos}},
  \bibinfo{author}{\bibfnamefont{M.}~\bibnamefont{Freitag}}, \bibnamefont{and}
  \bibinfo{author}{\bibfnamefont{P.}~\bibnamefont{Avouris}},
  \bibinfo{journal}{Phys. Rev. B} \textbf{\bibinfo{volume}{80}},
  \bibinfo{pages}{235402} (\bibinfo{year}{2009}).

\bibitem[{\citenamefont{Taniguchi and Watanabe}(2007)}]{Taniguchi_BN}
\bibinfo{author}{\bibfnamefont{T.}~\bibnamefont{Taniguchi}} \bibnamefont{and}
  \bibinfo{author}{\bibfnamefont{K.}~\bibnamefont{Watanabe}},
  \bibinfo{journal}{J. Cryst. Growth} \textbf{\bibinfo{volume}{303}},
  \bibinfo{pages}{525} (\bibinfo{year}{2007}).

\bibitem[{\citenamefont{Dean et~al.}(2010)\citenamefont{Dean, Young, Meric,
  Lee, Wang, Sorgenfrei, Watanabe, Taniguchi, Kim, Shepard et~al.}}]{Dean_BN}
\bibinfo{author}{\bibfnamefont{C.~R.} \bibnamefont{Dean}},
  \bibinfo{author}{\bibfnamefont{A.~F.} \bibnamefont{Young}},
  \bibinfo{author}{\bibfnamefont{I.}~\bibnamefont{Meric}},
  \bibinfo{author}{\bibfnamefont{C.}~\bibnamefont{Lee}},
  \bibinfo{author}{\bibfnamefont{L.}~\bibnamefont{Wang}},
  \bibinfo{author}{\bibfnamefont{S.}~\bibnamefont{Sorgenfrei}},
  \bibinfo{author}{\bibfnamefont{K.}~\bibnamefont{Watanabe}},
  \bibinfo{author}{\bibfnamefont{T.}~\bibnamefont{Taniguchi}},
  \bibinfo{author}{\bibfnamefont{P.}~\bibnamefont{Kim}},
  \bibinfo{author}{\bibfnamefont{K.~L.} \bibnamefont{Shepard}},
  \bibnamefont{et~al.}, \bibinfo{journal}{Nature Nanotech.}
  \textbf{\bibinfo{volume}{5}}, \bibinfo{pages}{722} (\bibinfo{year}{2010}).

\bibitem[{\citenamefont{Bao et~al.}(2010)\citenamefont{Bao, Zhao, Zhang, Liu,
  Kratz, Jing, Velasco, Smirnov, and Lau}}]{Bao_suspendedTri}
\bibinfo{author}{\bibfnamefont{W.}~\bibnamefont{Bao}},
  \bibinfo{author}{\bibfnamefont{Z.}~\bibnamefont{Zhao}},
  \bibinfo{author}{\bibfnamefont{H.}~\bibnamefont{Zhang}},
  \bibinfo{author}{\bibfnamefont{G.}~\bibnamefont{Liu}},
  \bibinfo{author}{\bibfnamefont{P.}~\bibnamefont{Kratz}},
  \bibinfo{author}{\bibfnamefont{L.}~\bibnamefont{Jing}},
  \bibinfo{author}{\bibfnamefont{J.}~\bibnamefont{Velasco}},
  \bibinfo{author}{\bibfnamefont{D.}~\bibnamefont{Smirnov}}, \bibnamefont{and}
  \bibinfo{author}{\bibfnamefont{C.~N.} \bibnamefont{Lau}},
  \bibinfo{journal}{Phys. Rev. Lett.} \textbf{\bibinfo{volume}{105}},
  \bibinfo{pages}{246601} (\bibinfo{year}{2010}).

\bibitem[{\citenamefont{Moser et~al.}(2007)\citenamefont{Moser, Barreiro, and
  Bachtold}}]{Moser_CurrentAnneal}
\bibinfo{author}{\bibfnamefont{J.}~\bibnamefont{Moser}},
  \bibinfo{author}{\bibfnamefont{A.}~\bibnamefont{Barreiro}}, \bibnamefont{and}
  \bibinfo{author}{\bibfnamefont{A.}~\bibnamefont{Bachtold}},
  \bibinfo{journal}{Appl. Phys. Lett.} \textbf{\bibinfo{volume}{91}},
  \bibinfo{pages}{163513} (\bibinfo{year}{2007}).

\bibitem[{\citenamefont{Du et~al.}(2008)\citenamefont{Du, Skachko, Barker, and
  Andrei}}]{Du_suspended}
\bibinfo{author}{\bibfnamefont{X.}~\bibnamefont{Du}},
  \bibinfo{author}{\bibfnamefont{I.}~\bibnamefont{Skachko}},
  \bibinfo{author}{\bibfnamefont{A.}~\bibnamefont{Barker}}, \bibnamefont{and}
  \bibinfo{author}{\bibfnamefont{E.~Y.} \bibnamefont{Andrei}},
  \bibinfo{journal}{Nature Nano.} \textbf{\bibinfo{volume}{3}},
  \bibinfo{pages}{491} (\bibinfo{year}{2008}).

\bibitem[{\citenamefont{Bolotin et~al.}(2008)\citenamefont{Bolotin, Sikes,
  Hone, Stormer, and Kim}}]{Bolotin_Tdepend}
\bibinfo{author}{\bibfnamefont{K.~I.} \bibnamefont{Bolotin}},
  \bibinfo{author}{\bibfnamefont{K.~J.} \bibnamefont{Sikes}},
  \bibinfo{author}{\bibfnamefont{J.}~\bibnamefont{Hone}},
  \bibinfo{author}{\bibfnamefont{H.~L.} \bibnamefont{Stormer}},
  \bibnamefont{and} \bibinfo{author}{\bibfnamefont{P.}~\bibnamefont{Kim}},
  \bibinfo{journal}{Phys. Rev. Lett.} \textbf{\bibinfo{volume}{101}},
  \bibinfo{pages}{096802} (\bibinfo{year}{2008}).

\bibitem[{\citenamefont{Piazza et~al.}(1999)\citenamefont{Piazza, Pellegrini,
  Beltram, Wegscheider, Jungwirth, and MacDonald}}]{Piazza_PhaseTranQHF}
\bibinfo{author}{\bibfnamefont{V.}~\bibnamefont{Piazza}},
  \bibinfo{author}{\bibfnamefont{V.}~\bibnamefont{Pellegrini}},
  \bibinfo{author}{\bibfnamefont{F.}~\bibnamefont{Beltram}},
  \bibinfo{author}{\bibfnamefont{W.}~\bibnamefont{Wegscheider}},
  \bibinfo{author}{\bibfnamefont{T.}~\bibnamefont{Jungwirth}},
  \bibnamefont{and} \bibinfo{author}{\bibfnamefont{A.~H.}
  \bibnamefont{MacDonald}}, \bibinfo{journal}{Nature}
  \textbf{\bibinfo{volume}{402}}, \bibinfo{pages}{638} (\bibinfo{year}{1999}).

\bibitem[{\citenamefont{Zhang et~al.}(2005{\natexlab{b}})\citenamefont{Zhang,
  Faulhaber, and Jiang}}]{Zhang_LLcrossingTwo-subband}
\bibinfo{author}{\bibfnamefont{X.~C.} \bibnamefont{Zhang}},
  \bibinfo{author}{\bibfnamefont{D.~R.} \bibnamefont{Faulhaber}},
  \bibnamefont{and} \bibinfo{author}{\bibfnamefont{H.~W.} \bibnamefont{Jiang}},
  \bibinfo{journal}{Phys. Rev. Lett.} \textbf{\bibinfo{volume}{95}},
  \bibinfo{pages}{216801} (\bibinfo{year}{2005}{\natexlab{b}}).

\bibitem[{\citenamefont{Dresselhaus and
  Dresselhaus}(2002)}]{Dresselhaus_Graphite}
\bibinfo{author}{\bibfnamefont{M.~S.} \bibnamefont{Dresselhaus}}
  \bibnamefont{and}
  \bibinfo{author}{\bibfnamefont{G.}~\bibnamefont{Dresselhaus}},
  \bibinfo{journal}{Adv. Phys.} \textbf{\bibinfo{volume}{51}},
  \bibinfo{pages}{1} (\bibinfo{year}{2002}).

\bibitem[{\citenamefont{Koshino and McCann}(2009)}]{Mikito_ABC}
\bibinfo{author}{\bibfnamefont{M.}~\bibnamefont{Koshino}} \bibnamefont{and}
  \bibinfo{author}{\bibfnamefont{E.}~\bibnamefont{McCann}},
  \bibinfo{journal}{Phys. Rev. B} \textbf{\bibinfo{volume}{80}},
  \bibinfo{pages}{165409} (\bibinfo{year}{2009}).

\bibitem[{\citenamefont{Zhang et~al.}(2006)\citenamefont{Zhang, Jiang, Small,
  Purewal, Tan, Fazlollahi, Chudow, Jaszczak, Stormer, and
  Kim}}]{Zhang_MonoLLSplitting}
\bibinfo{author}{\bibfnamefont{Y.}~\bibnamefont{Zhang}},
  \bibinfo{author}{\bibfnamefont{Z.}~\bibnamefont{Jiang}},
  \bibinfo{author}{\bibfnamefont{J.~P.} \bibnamefont{Small}},
  \bibinfo{author}{\bibfnamefont{M.~S.} \bibnamefont{Purewal}},
  \bibinfo{author}{\bibfnamefont{Y.-W.} \bibnamefont{Tan}},
  \bibinfo{author}{\bibfnamefont{M.}~\bibnamefont{Fazlollahi}},
  \bibinfo{author}{\bibfnamefont{J.~D.} \bibnamefont{Chudow}},
  \bibinfo{author}{\bibfnamefont{J.~A.} \bibnamefont{Jaszczak}},
  \bibinfo{author}{\bibfnamefont{H.~L.} \bibnamefont{Stormer}},
  \bibnamefont{and} \bibinfo{author}{\bibfnamefont{P.}~\bibnamefont{Kim}},
  \bibinfo{journal}{Phys. Rev. Lett.} \textbf{\bibinfo{volume}{96}},
  \bibinfo{pages}{136806} (\bibinfo{year}{2006}).

\bibitem[{\citenamefont{Checkelsky et~al.}(2008)\citenamefont{Checkelsky, Li,
  and Ong}}]{Checkelsky_nuZero}
\bibinfo{author}{\bibfnamefont{J.~G.} \bibnamefont{Checkelsky}},
  \bibinfo{author}{\bibfnamefont{L.}~\bibnamefont{Li}}, \bibnamefont{and}
  \bibinfo{author}{\bibfnamefont{N.~P.} \bibnamefont{Ong}},
  \bibinfo{journal}{Phys. Rev. Lett.} \textbf{\bibinfo{volume}{100}},
  \bibinfo{pages}{206801} (\bibinfo{year}{2008}).

\bibitem[{\citenamefont{Feldman et~al.}(2009)\citenamefont{Feldman, Martin, and
  Yacoby}}]{Feldman_bilayer}
\bibinfo{author}{\bibfnamefont{B.~E.} \bibnamefont{Feldman}},
  \bibinfo{author}{\bibfnamefont{J.}~\bibnamefont{Martin}}, \bibnamefont{and}
  \bibinfo{author}{\bibfnamefont{A.}~\bibnamefont{Yacoby}},
  \bibinfo{journal}{Nature Phys.} \textbf{\bibinfo{volume}{5}},
  \bibinfo{pages}{889} (\bibinfo{year}{2009}).

\bibitem[{\citenamefont{Zhao et~al.}(2009)\citenamefont{Zhao, Cadden-Zimansky,
  Jiang, and Kim}}]{Zhao_SymBreakBi}
\bibinfo{author}{\bibfnamefont{Y.}~\bibnamefont{Zhao}},
  \bibinfo{author}{\bibfnamefont{P.}~\bibnamefont{Cadden-Zimansky}},
  \bibinfo{author}{\bibfnamefont{Z.}~\bibnamefont{Jiang}}, \bibnamefont{and}
  \bibinfo{author}{\bibfnamefont{P.}~\bibnamefont{Kim}},
  \bibinfo{journal}{Phys. Rev. Lett.} \textbf{\bibinfo{volume}{104}},
  \bibinfo{pages}{066801} (\bibinfo{year}{2009}).

\bibitem[{\citenamefont{Du et~al.}(2009)\citenamefont{Du, Skachko, Duerr,
  Luican, and Andrei}}]{Du_fraction}
\bibinfo{author}{\bibfnamefont{X.}~\bibnamefont{Du}},
  \bibinfo{author}{\bibfnamefont{I.}~\bibnamefont{Skachko}},
  \bibinfo{author}{\bibfnamefont{F.}~\bibnamefont{Duerr}},
  \bibinfo{author}{\bibfnamefont{A.}~\bibnamefont{Luican}}, \bibnamefont{and}
  \bibinfo{author}{\bibfnamefont{E.~Y.} \bibnamefont{Andrei}},
  \bibinfo{journal}{Nature} \textbf{\bibinfo{volume}{462}},
  \bibinfo{pages}{192} (\bibinfo{year}{2009}).

\bibitem[{\citenamefont{Song et~al.}(2010)\citenamefont{Song, Otte, Kuk, Hu,
  Torrance, First, de~Heer, Min, Adam, Stiles et~al.}}]{Song_N1splitting}
\bibinfo{author}{\bibfnamefont{Y.~J.} \bibnamefont{Song}},
  \bibinfo{author}{\bibfnamefont{A.~F.} \bibnamefont{Otte}},
  \bibinfo{author}{\bibfnamefont{Y.}~\bibnamefont{Kuk}},
  \bibinfo{author}{\bibfnamefont{Y.}~\bibnamefont{Hu}},
  \bibinfo{author}{\bibfnamefont{D.~B.} \bibnamefont{Torrance}},
  \bibinfo{author}{\bibfnamefont{P.~N.} \bibnamefont{First}},
  \bibinfo{author}{\bibfnamefont{W.~A.} \bibnamefont{de~Heer}},
  \bibinfo{author}{\bibfnamefont{H.}~\bibnamefont{Min}},
  \bibinfo{author}{\bibfnamefont{S.}~\bibnamefont{Adam}},
  \bibinfo{author}{\bibfnamefont{M.~D.} \bibnamefont{Stiles}},
  \bibnamefont{et~al.}, \bibinfo{journal}{Nature}
  \textbf{\bibinfo{volume}{467}}, \bibinfo{pages}{185} (\bibinfo{year}{2010}).

\bibitem[{\citenamefont{Jungwirth et~al.}(1998)\citenamefont{Jungwirth, Shukla,
  Smr\ifmmode~\check{c}\else \v{c}\fi{}ka, Shayegan, and
  MacDonald}}]{Jungwirth_QHF}
\bibinfo{author}{\bibfnamefont{T.}~\bibnamefont{Jungwirth}},
  \bibinfo{author}{\bibfnamefont{S.~P.} \bibnamefont{Shukla}},
  \bibinfo{author}{\bibfnamefont{L.}~\bibnamefont{Smr\ifmmode~\check{c}\else
  \v{c}\fi{}ka}}, \bibinfo{author}{\bibfnamefont{M.}~\bibnamefont{Shayegan}},
  \bibnamefont{and} \bibinfo{author}{\bibfnamefont{A.~H.}
  \bibnamefont{MacDonald}}, \bibinfo{journal}{Phys. Rev. Lett.}
  \textbf{\bibinfo{volume}{81}}, \bibinfo{pages}{2328} (\bibinfo{year}{1998}).

\bibitem[{\citenamefont{Ezawa}(2007)}]{Motohiko_QHEtrilayer}
\bibinfo{author}{\bibfnamefont{M.}~\bibnamefont{Ezawa}},
  \bibinfo{journal}{Physica E} \textbf{\bibinfo{volume}{40}},
  \bibinfo{pages}{269} (\bibinfo{year}{2007}).

\bibitem[{\citenamefont{Koshino and McCann}(2010)}]{Mikito_ParityValley}
\bibinfo{author}{\bibfnamefont{M.}~\bibnamefont{Koshino}} \bibnamefont{and}
  \bibinfo{author}{\bibfnamefont{E.}~\bibnamefont{McCann}},
  \bibinfo{journal}{Phys. Rev. B} \textbf{\bibinfo{volume}{81}},
  \bibinfo{pages}{115315} (\bibinfo{year}{2010}).

\end{thebibliography}
\end{document}